\documentclass{elsart}
\usepackage[dvips]{graphicx}
\newcommand{\krz}{\ensuremath{K^{*0}}}
\newcommand{\krzb}{\ensuremath{\overline{K}^{*0}}}
\newcommand{\krzmndk}{\ensuremath{D^+ \rightarrow \krzb \mu^+ \nu}}
\newcommand{\krzlndk}{\ensuremath{D^+ \rightarrow \krzb \ell^+ \nu_\ell}}

\newcommand{\kpimndk}{\ensuremath{D^+ \rightarrow K^- \pi^+ \mu^+ \nu }}

\newcommand{\gevcsq}{\ensuremath{\textrm{GeV}/c^2}}

\newcommand{\thv}{\ensuremath{\theta_\textrm{v}}}
\newcommand{\thl}{\ensuremath{\theta_\ell}}
\newcommand{\costhv}{\ensuremath{\cos\thv}}

\newcommand{\costhl}{\ensuremath{\cos\thl}}

\newcommand{\qsq}{\ensuremath{q^2}}
\newcommand{\yvar}{\ensuremath{\qsq{}/\qsq_{\rm max}}}
\newcommand{\bw}{\ensuremath{\textrm{B}_{\krz}}}
\newcommand{\mkpi}{\ensuremath{m_{K\pi}}}

\newcommand{\gevc}{\ensuremath{\textrm{GeV}^2/c^2}}

\newcommand{\rvvalue}{\ensuremath{1.504 \pm 0.057 \pm 0.039}}
\newcommand{\rtwovalue}{\ensuremath{0.875 \pm 0.049 \pm 0.064}}
\newcommand{\rvresult}{\ensuremath{r_v = \rvvalue{}}}
\newcommand{\rtworesult}{\ensuremath{r_2 = \rtwovalue{}}}
\newcommand{\rtwo}{\ensuremath{r_2}}
\newcommand{\rthree}{\ensuremath{r_3}}
\newcommand{\rvee}{\ensuremath{r_v}}
\newcommand{\mysection}[1]{\section{#1}}

\newcounter{saveeqn}%

\begin{document}
\begin{frontmatter}
\title{New Measurements of the \krzmndk{} Form Factor Ratios}
The FOCUS Collaboration%
\footnote{See \textrm{http://www-focus.fnal.gov/authors.html} for
additional author information.}
\author[ucd]{J.~M.~Link}
\author[ucd]{M.~Reyes}
\author[ucd]{P.~M.~Yager}
\author[cbpf]{J.~C.~Anjos}
\author[cbpf]{I.~Bediaga}
\author[cbpf]{C.~G\"obel}
\author[cbpf]{J.~Magnin}
\author[cbpf]{A.~Massafferri}
\author[cbpf]{J.~M.~de~Miranda}
\author[cbpf]{I.~M.~Pepe}
% ...

\address[ucd]{University of California, Davis, CA 95616}
\address[cbpf]{Centro Brasileiro de Pesquisas F\'isicas, Rio de Janeiro, RJ, Brasil}
\address[cinv]{CINVESTAV, 07000 M\'exico City, DF, Mexico}
\address[cu]{University of Colorado, Boulder, CO 80309}
\address[fnal]{Fermi National Accelerator Laboratory, Batavia, IL 60510}
\address[fras]{Laboratori Nazionali di Frascati dell'INFN, Frascati, Italy I-00044}
\address[ui]{University of Illinois, Urbana-Champaign, IL 61801}
\address[iu]{Indiana University, Bloomington, IN 47405}
\address[korea]{Korea University, Seoul, Korea 136-701}
\address[kp]{Kyungpook National University, Taegu, Korea 702-701}
\address[milan]{INFN and University of Milano, Milano, Italy}
\address[nc]{University of North Carolina, Asheville, NC 28804}
\address[pavia]{Dipartimento di Fisica Nucleare e Teorica and INFN, Pavia, Italy}
\address[pr]{University of Puerto Rico, Mayaguez, PR 00681}
\address[sc]{University of South Carolina, Columbia, SC 29208}
\address[ut]{University of Tennessee, Knoxville, TN 37996}
\address[vu]{Vanderbilt University, Nashville, TN 37235}
\address[wisc]{University of Wisconsin, Madison, WI 53706}

\nobreak
\begin{abstract}
Using a large sample of $D^+ \rightarrow K^- \pi^+ \mu^+ \nu$ decays
collected by the FOCUS photoproduction experiment at Fermilab, we
present new measurements of two semileptonic form factor ratios: $r_v$
and $r_2$.  We find \rvresult{} and \rtworesult{}.  Our form factor
results include the effects of the $s$-wave interference discussed in
Reference~\cite{anomaly}.
\end{abstract}
\end{frontmatter}
\newpage
\newpage

\mysection{Introduction}

This paper provides new measurements of the parameters that
describe \kpimndk{} decay.  In an earlier paper~\cite{anomaly} we 
described this process as the interference of a \krzmndk{} amplitude with
a constant $s$-wave amplitude.  The \krzmndk{} decay amplitude
is described~\cite{KS} by four form factors with an assumed (pole form) \qsq{}
dependence.  Following earlier experimental work~
\cite{beatrice,e791e,e791mu,e687,e653,e691},
the \krzmndk{} amplitude is then described by ratios of form factors taken
at \qsq{} = 0. The traditional set is: \rtwo{}, \rthree{}, and \rvee{}
which we define explicitly after Equation \ref{amp1}.

\begin{figure}[tbph!]
 \begin{center}
  \includegraphics[width=3.0in]{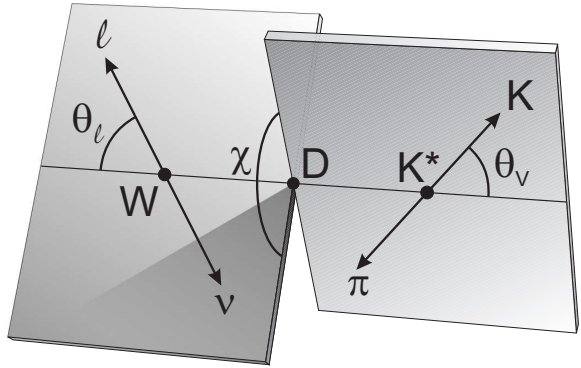}
  \caption{Definition of kinematic variables.
 \label{angles}}
 \end{center}
\end{figure}

Five kinematic variables that uniquely describe \kpimndk{} decay are
illustrated in Figure~\ref{angles}. These are the $K^- \pi^+$
invariant mass (\mkpi{}) , the square of the $\mu\nu$ mass (\qsq{}),
and three decay angles: the angle between the $\pi$ and the $D$
direction in the $K^- \pi^+$ rest frame (\thv{}), the angle between
the $\nu$ and the $D$ direction in the $\mu\nu$ rest frame (\thl{}),
and the acoplanarity angle between the two decay planes
($\chi$). These angular conventions on \thl{} and \thv{} apply to
both the $D^+$ and $D^-$. The sense of the acoplanarity variable is
defined via a cross product expression of the form: $ (\vec P_\mu
\times \vec P_\nu) \times (\vec P_K \times \vec P_\pi) \cdot \vec P_{K
\pi}$ where all momentum vectors are in the $D^+$ rest frame.  Since
this expression involves five momentum vectors, as one goes from $D^+
\rightarrow D^-$ one must change $\chi \rightarrow -\chi$ in
Equation~\ref{amp1} to get the same intensity for the $D^+$ and $D^-$ 
assuming \emph{CP} symmetry.

Using
the notation of \cite{KS}, we write the decay distribution for
\kpimndk{} in terms of the four helicity basis form factors:
$H_+~,~H_0~,~H_-~,H_t$.  
\newpage
\begin{eqnarray}
{d^5 \Gamma \over dm_{K \pi}~d\qsq~d\cos\thv~d\cos\thl~d\chi}
\propto K (\qsq{} - m_l^2 )\left\{\left| \begin{array}{l}
 (1 + \cos \thl )\sin \thv e^{i\chi } \bw H_ +   \\
  - \,(1 - \cos \thl )\sin \thv e^{-i\chi } \bw H_ -   \\
  - \,2\sin \thl (\cos \thv \bw + Ae^{i\delta } )H_0  \\
 \end{array} \right|^2 \right. \nonumber \\
 + \frac{m^2_\ell}{\qsq} \left. \left| \begin{array}{l}
 \sin\thl \sin\thv \bw \left(e^{i\chi}H_+ + e^{-i\chi}H_-\right) \\
+\,2\cos\thl\left(\cos\thv\bw + Ae^{i\delta}\right)H_0 \\
+\,2\left(cos\thv\bw + Ae^{i\delta}\right)H_t
 \end{array} \right|^2 \right\}\phantom{xxx}
\label{amp1}
\end{eqnarray}

where $K$ is the momentum of the $K^- \pi^+$ system in the rest frame
of the $D^+$. The first term gives the intensity for the $\mu^+$ to be
right-handed, while the (highly suppressed) second term gives the
intensity for it to be left-handed.  The helicity basis form factors
are given by:
\begin{eqnarray*}
H_\pm (\qsq) &=& (M_D+\mkpi)A_1(\qsq)\mp 2{M_D K\over M_D+m_{K\pi}}V(\qsq)\\
H_0 (\qsq) &=& {1\over 2\mkpi\sqrt{\qsq}}
\left[ (M^2_D -m^2_{K\pi}-\qsq)(M_D+\mkpi)A_1(\qsq)
-4{M^2_D K^2\over M_D+\mkpi}A_2(\qsq) \right] \\
H_t(\qsq) &=& {M_D K\over M_{K\pi}\sqrt{\qsq}}
\left[  (M_D+M_{K\pi})A_1(\qsq) - {(M^2_D -M^2_{K\pi}+\qsq) \over M_D+M_{K\pi}}A_2(\qsq)
+{2\qsq\over M_D+M_{K\pi}}A_3(\qsq) \right]
\end{eqnarray*}
The vector and axial form factors are generally parameterized by a pole
dominance form:
\[
A_i(\qsq)={A_i(0)\over 1-\qsq/M_A^2}~~~~~~~~
V(\qsq)={V(0)\over 1-\qsq/M_V^2}
\]
where we use nominal (spectroscopic) pole masses of $M_A =
2.5~\gevcsq$ and $M_V = 2.1~\gevcsq$.

\footnote{Equation \ref{amp1} implicitly assumes that the \qsq{}
dependence of the $s$-wave amplitude coupling to the virtual $W^+$ is
the same as the $H_0$ form factor describing the \krzmndk{}, but there
is no theoretical justification for this assumption. This $q^2$
dependence is compared to the data in Reference~\cite{anomaly}.  We
tried form factor ratio fits with an alternative, significantly
different kinematic dependence for the $s$-wave amplitude where
$H_0(q^2)$ is replaced by $K/(1-\qsq{}/M_A^2)$ in Equation \ref{amp1}.
We found that the values of $r_2$ and $r_v$ changed by less than 6\%
of their statistical error when fit with this alternative form.}  The
\bw{} denotes the Breit-Wigner amplitude describing the \krzb{}
resonance:\footnote {We are using a $p$-wave Breit-Wigner form with a
width proportional to the cube of the kaon momentum in the kaon-pion
rest frame ($P^*$) over the value of this momentum when the kaon-pion
mass equals the resonant mass ($P^*_0$).  The squared modulus of our
Breit-Wigner form will have an effective $P^{*3}$ dependence in the
numerator as well. Two powers $P^*$ come explicitly from the $P^*$ in
the numerator of the amplitude and one power arises from the 4 body
phase space.}
\[
\bw = \frac{\sqrt{m_0 \Gamma\,} \left(\frac{P^*}{P_0^*}\right)}
	   {m_{K\pi}^2 - m_0^2 + i m_0 \Gamma 
	\left(\frac{P^*}{P_0^*}\right)^3} 
\]
Under these assumptions, the decay
intensity is then parameterized by the $\rvee{} \equiv V(0)/ A_1(0) ~,~
\rtwo{} \equiv A_2(0)/A_1(0)~,~\rthree{} \equiv A_3(0)/A_1(0)$ form factor 
ratios describing the \krzmndk{} amplitude and the modulus $A$ and phase
$\delta$ describing the $s$-wave amplitude.
Throughout this paper, unless explicitly stated otherwise,
the charge conjugate is also implied when a decay mode of a specific
charge is stated.

\mysection{Experimental and analysis details}

The data for this paper were collected in the Wideband photoproduction
experiment FOCUS during the Fermilab 1996--1997 fixed-target run. In
FOCUS, a forward multi-particle spectrometer is used to measure the
interactions of high energy photons on a segmented BeO target. The
FOCUS detector is a large aperture, fixed-target spectrometer with
excellent vertexing and particle identification. Most of the FOCUS
experiment and analysis techniques have been described
previously~\cite{anomaly,nim,ycp,CNIM}.
Our analysis cuts were chosen to give reasonably uniform acceptance
over the five kinematic decay variables, while maintaining a strong
rejection of backgrounds.  To suppress background from the
re-interaction of particles in the target region which can mimic a
decay vertex, we required that the charm secondary vertex was located
at least one  standard deviation outside of all solid
material including our target and target microstrip system.

To isolate the
\kpimndk{} topology, we required that candidate muon, pion, and kaon
tracks appeared in a secondary vertex with a confidence level
exceeding 5\%.  The muon track, when extrapolated to the shielded muon
arrays, was required to match muon hits with a confidence level
exceeding 5\%. The kaon was required to have a \v Cerenkov light
pattern more consistent with that for a kaon than that for a pion by 1
unit of log likelihood, while the pion track was
required to have a light pattern favoring the pion hypothesis over
that for the kaon by 1 unit~\cite{CNIM}.

To further reduce muon misidentification, a muon candidate was allowed
to have at most one missing hit in the 6 planes comprising our inner
muon system and an energy exceeding 10 GeV.  In order to suppress
muons from pions and kaons decaying within our apparatus, we required
that each muon candidate had a confidence level exceeding 2\% to the
hypothesis that it had a consistent trajectory through our two
analysis magnets.

Non-charm and random combinatoric
backgrounds were reduced by requiring both a detachment between the
vertex containing the $K^-\pi^+\mu^+$ and the primary production
vertex of 10 standard deviations and a minimum visible energy
$(E_K+E_\pi+E_\mu)$ of 30 GeV. To suppress possible backgrounds from
higher multiplicity charm decay, we isolate the $K\pi\mu$ vertex from
other tracks in the event (not including tracks in the primary vertex)
by requiring that the maximum confidence level for another track to
form a vertex with the candidate be less than 0.1\%.

\begin{figure}[tbph!]
 \begin{center}
  \includegraphics[width=3.5in]{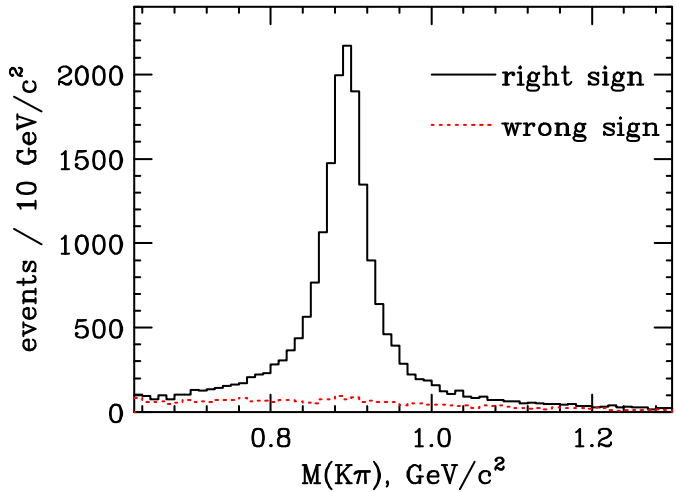}
  \caption{\kpimndk{} signal.  Right-sign and
  wrong-sign samples are shown. In the mass window from 0.8--1.0 \gevcsq{}
  there is a right-sign excess of 14\,678 events.  A Monte Carlo that
  simulates the production and decay of all known charm species
  predicts that $\approx$3\% of this excess is actually background
  from other charm decays.
\label{signal}} \end{center}
\end{figure}

In order to allow for the missing energy of the neutrino in this
semileptonic $D^+$ decay, we required the reconstructed $K \pi \mu$
mass be less than the nominal $D^+$ mass.  Background from $D^+
\rightarrow K^- \pi^+ \pi^+$, where a pion is misidentified as a muon,
was reduced using a mass cut: we required when the muon track is
treated as a pion and the combination is reconstructed as a $K \pi
\pi$, the $K \pi \pi$ invariant mass differed from the nominal $D^+$
mass by at least three standard deviations.  In order to suppress
background from $D^{*+} \rightarrow D^0 \pi^+ \rightarrow (K^- \mu^+
\nu) \pi^+$ we required $M(K^- \mu^+ \nu \pi^+) - M(K^- \mu^+ \nu) >
0.18~\gevcsq $. The \mkpi{} distribution for our \kpimndk{} candidates
is shown in Figure \ref{signal}.

The technique used to reconstruct the neutrino momentum through the
$D^+$ line-of-flight, and tests of our ability to simulate the
resolution on kinematic variables that rely on the neutrino momentum are
described in Reference~\cite{anomaly}.

\mysection{Fitting Technique} 

We use a binned version~\cite{will} of the fitting technique developed
by the E691 Collaboration~\cite{schmidt} for fitting decay intensities
where some of the kinematic variables have very poor resolution such
as the four variables that rely on reconstructed neutrino kinematics.
The observed number of wrong-sign-subtracted events in each kinematic
bin is compared to a prediction. The production is constructed from a
\kpimndk{} signal Monte Carlo incorporating $s$-wave interference~
\cite{anomaly} plus a wrong-sign-subtracted, charm background
contribution predicted by a background Monte Carlo which simulates all
known charm decays as well as our misidentification levels.

Although the charm background correction was fairly unimportant given
the tight muon cuts used for our quoted results, this correction was
important when looser muon cuts were employed.  In the sample selected
with looser muon cuts, the charm background increased from about 3\%
to 7\% of the total right-sign excess. In fits to the looser sample,
the charm background correction typically lowered the uncorrected
\rvee{} by 0.15 (or about 2.7 times our statistical error) to a value
very consistent with our quoted result.  The charm background is
primarily due to false muons from decays of pions and kaons in flight
and therefore tends to populate low (lab) momentum or the negative
\costhl{} region.  Including the background correction to the fit
reduced the apparent backward-forward asymmetry in \costhl{}, thus
reducing the difference between the $H_+$ and $H_-$ form factor in
Equation \ref{amp1}. Since this difference is proportional to \rvee{},
fits with the background correction will have a lower \rvee{} relative
to fits where no charm background correction was performed.  The
effect of charm backgrounds on the \rtwo{} form factor ratio was
found to be much smaller --- about $1 \sigma$ for our ``loose'' muon
fits.

The signal Monte Carlo was initially
generated flat in the $K \pi \mu \nu$ phase space and the five generated
as well as reconstructed kinematic variables were stored for each
event. The signal prediction for a given fit iteration is then
computed by weighting each event within a given reconstructed
kinematic bin by the intensity given by Equation \ref{amp1} evaluated
using the five generated kinematic variables for the current set of fit
parameters.  The background Monte Carlo was normalized 
to the observed
number of \kpimndk{} events in the mass range
$0.8 < \mkpi < 1.0~\gevcsq{}$ after applying the wrong-sign subtraction. 
The signal Monte Carlo was normalized
to the difference between the observed wrong-sign-subtracted yield and
the predicted wrong-sign-subtracted background yield.  The fit
determined the physics parameters by minimizing the $\chi^2$ over all
bins.

Two fits were employed in this analysis: a fit to the $s$-wave amplitude
with fixed \rvee{} and \rtwo{} form factor ratios, and a fit to the
\rvee{} and \rtwo{} form factors ratios\footnote{We decided not to fit for
the \rthree{} form factor ratio since our anticipated \rthree{} error
given our sample size and the \qsq{} cut described shortly would be
$\pm 3$.} with a fixed $s$-wave amplitude and phase. In the form
factor fit, we used five bins in \costhv{}, five bins in \costhl{},
three bins in $|\chi|$, and three bins in \yvar{} for a total of 225 bins.
This binning was chosen to be sensitive to the main features of our
model intensity, Equation \ref{amp1}, that depend on \rvee{} and
\rtwo{}. The $s$-wave amplitude used three bins of \costhv{} , three bins of
\costhl, four bins of \mkpi{}, and three bins of $\chi$ for a total of
108 bins.  The $s$-wave amplitude binning was chosen to emphasize the
\mkpi{} dependence of the angular distribution. This dependence is
extremely sensitive to the $s$-wave phase as discussed in
Reference~\cite{anomaly}.  In both cases, evenly spaced bins were used.
The binnings of both fits were chosen to ensure at least 10 observed
events per fit bin.  These two fits were very loosely coupled so only
a few iterations sufficed to obtain stable results.

\begin{figure}[htp]
\includegraphics[height=2.in]{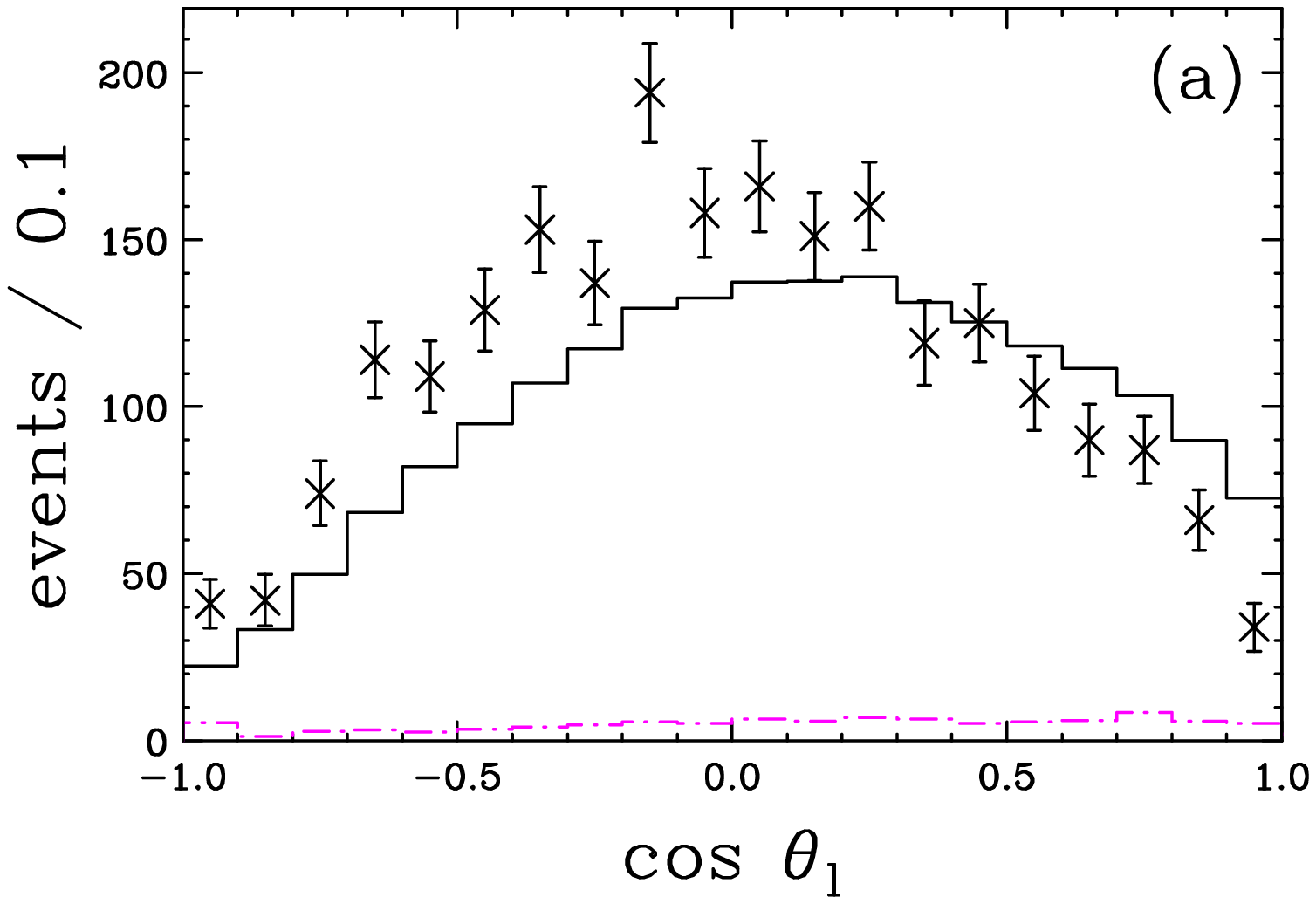}
\includegraphics[height=2.in]{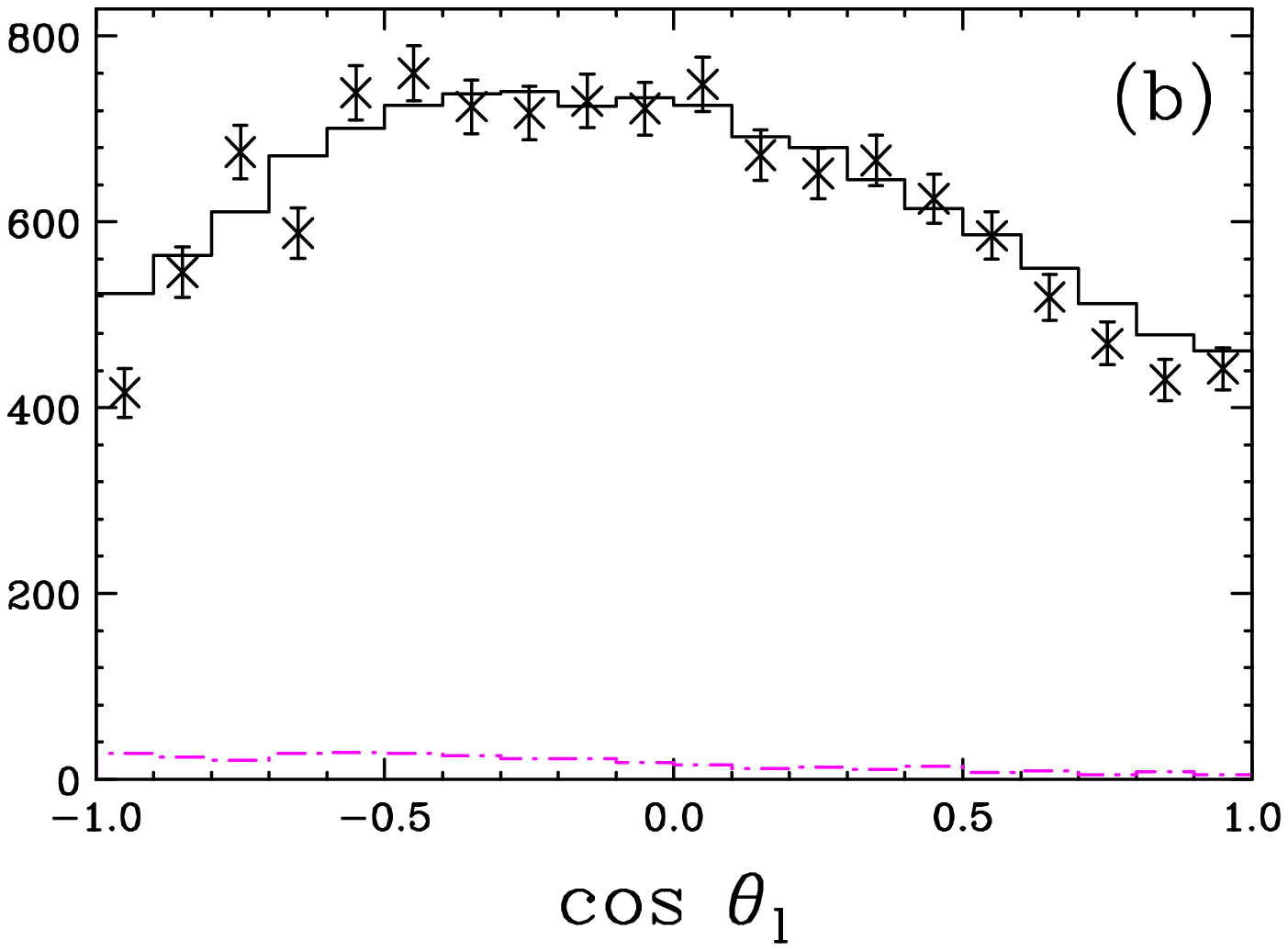}
\caption{We show the \costhl{} distribution in two ranges for
\qsq{}. The data are shown as points with error bars, 
while the Monte Carlo model which includes
charm backgrounds are the solid line histograms.
The predicted charm background contributions are
the dashed lines. Figure (a) shows events in the low \qsq{} region :
$\qsq{} < 0.2~\gevc{}$. Figure (b) shows events in the remaining \qsq{} region
: $\qsq{} > 0.2~\gevc{}$.  
\label{lowq_cosl}}
\end{figure}

Our initial form factor fits were of very poor quality due to a
problem with our model matching the \costhl{} distribution at very low
\qsq{} ($\qsq{} < 0.2~\gevc{}$).  Figure \ref{lowq_cosl} illustrates
this problem by comparing the \costhl{} distribution in data and our
model for events below \qsq{} = 0.2 \gevc{} and above \qsq{} = 0.2
\gevc{} where the discrepancy is far less. Excluding the $\qsq{} <
0.2~\gevc{}$ region caused the $\chi^2$ of our fits to reduce by 86
units. The low \qsq{} discrepancy can most easily be explained as a
deviation from the assumed pole dominance of the vector form factor,
$V(\qsq{})$, but we have not eliminated all other possibilities.  We
have decided to exclude this region from our form factor and $s$-wave
amplitude fits. When these regions were excluded, the fitted \rvee{}
and \rtwo{} form factor ratios decreased by 1.2 $\sigma$ and 0.4
$\sigma$ respectively.  With the $\qsq{}< 0.2~\gevc{}$ removed, our
form factor fit has a $\chi^2$ per degree of freedom of 1.15 for 223
degrees of freedom or a confidence level of 5.2\%.

\begin{figure}[htp]
\includegraphics[height=2.in]{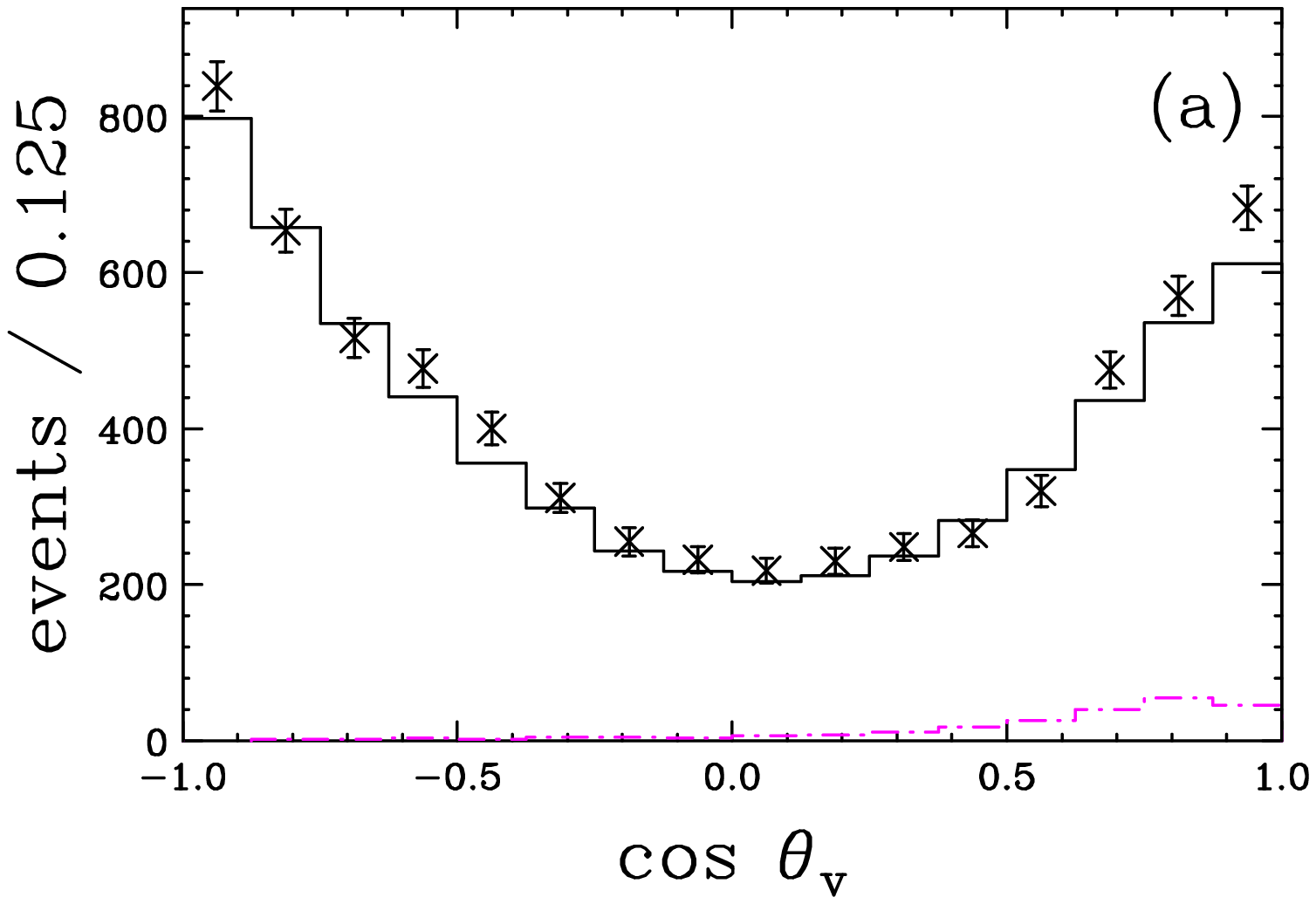}
\includegraphics[height=2.in]{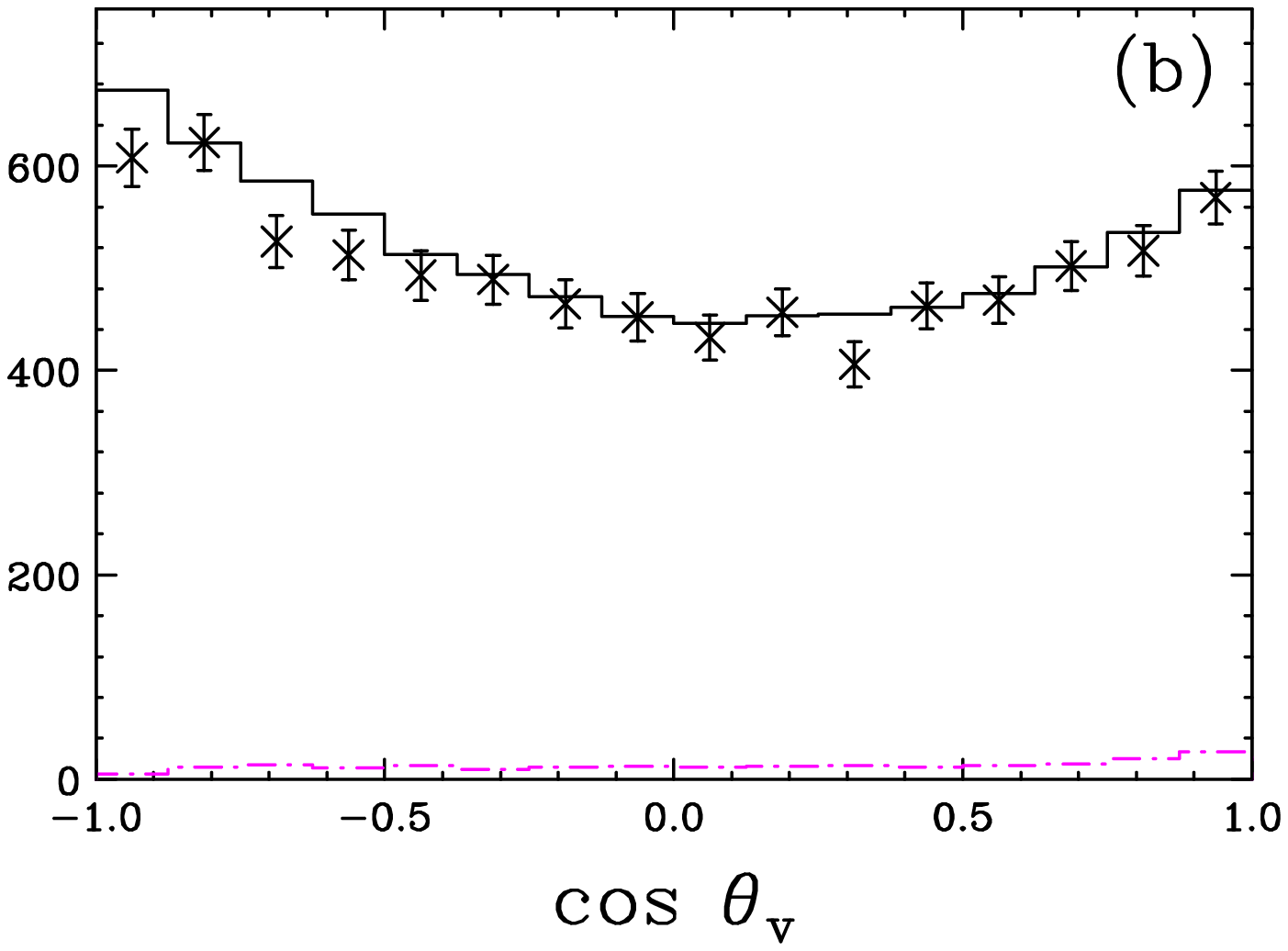}
\includegraphics[height=2.in]{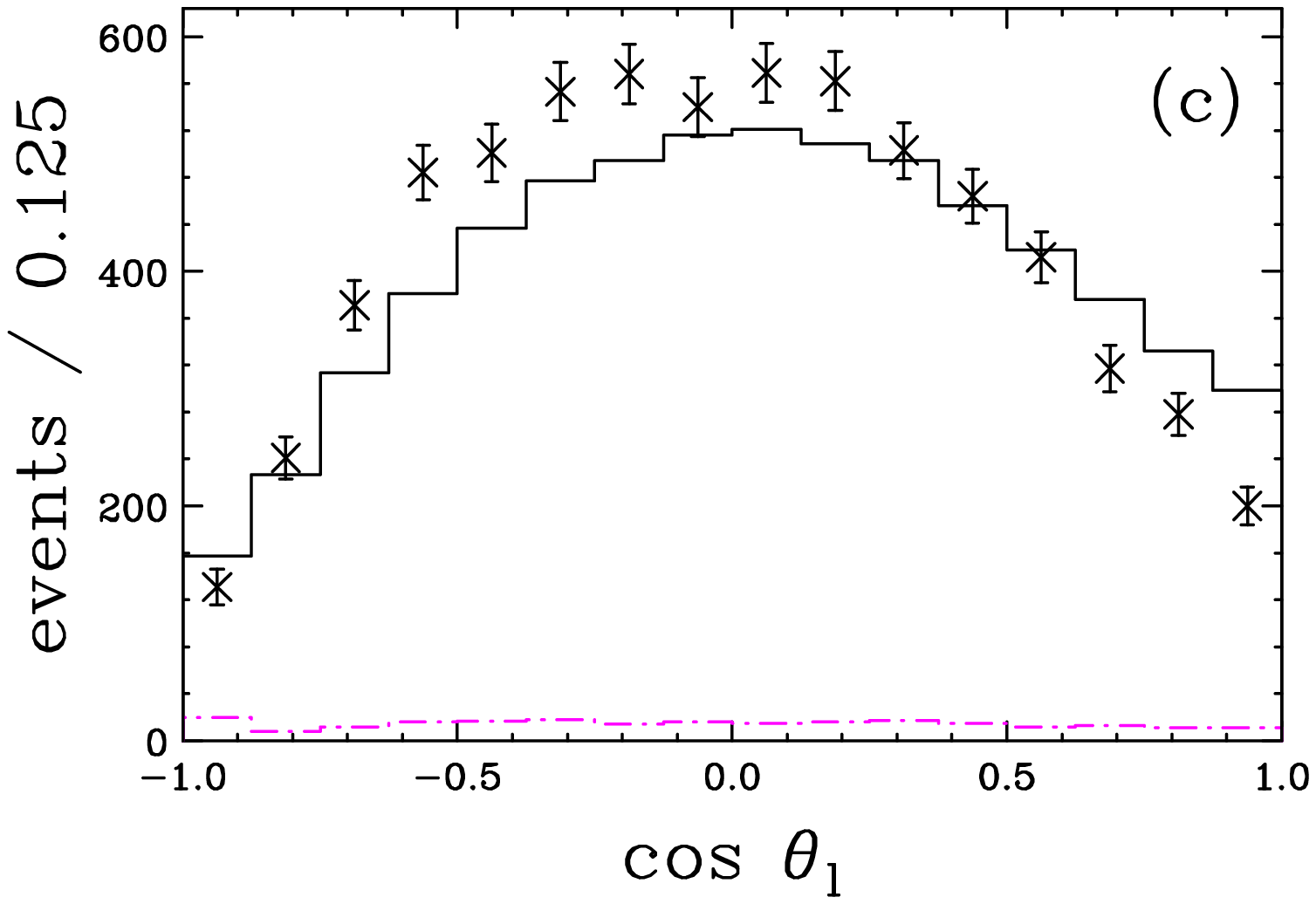}
\includegraphics[height=2.in]{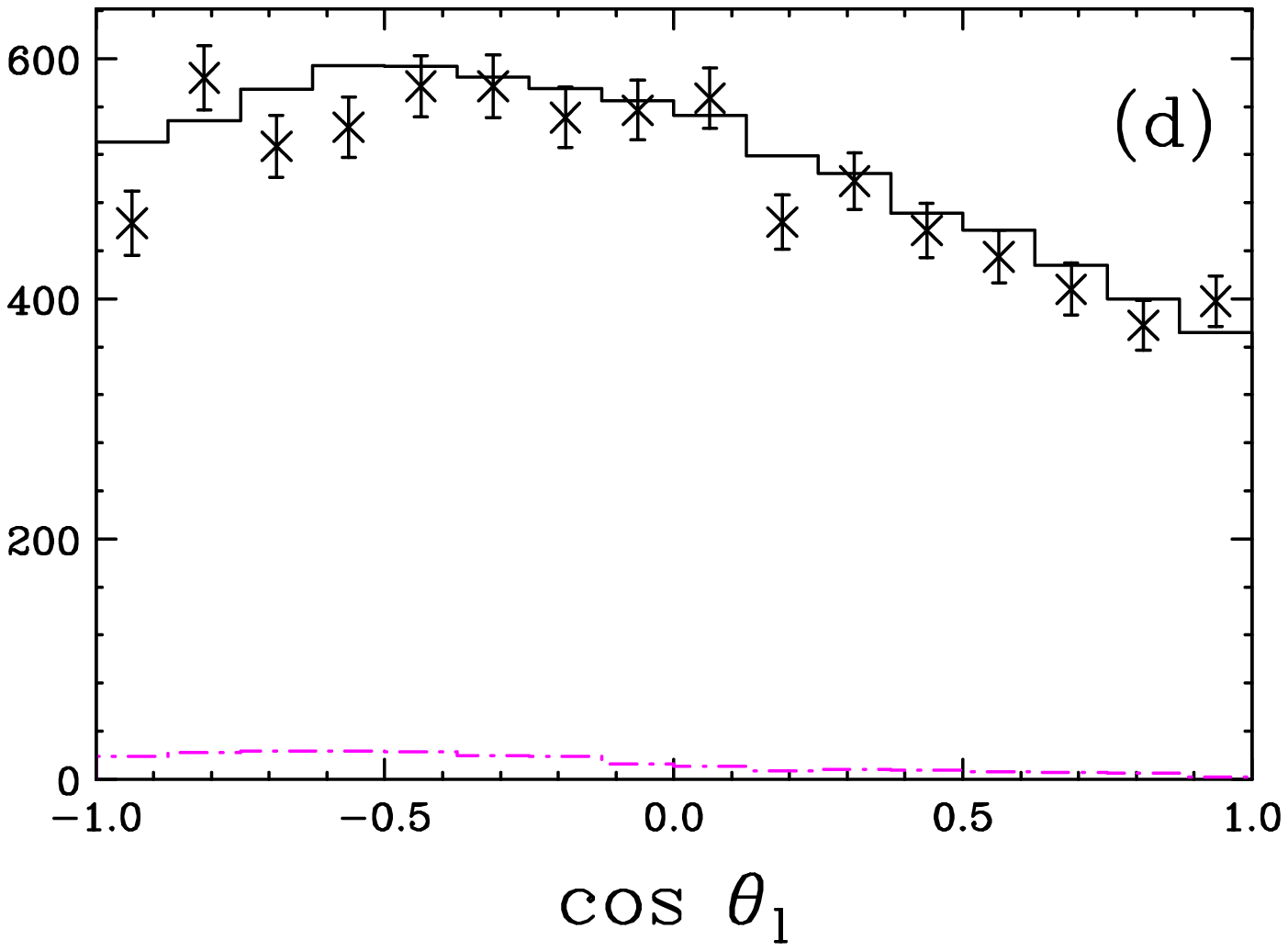}
\caption{Various \costhv{} and \costhl{} projections.  The data are
the points with error bars. The MC model predictions are the solid
line histograms. The predicted charm background projections are the
dashed lines.  The model distributions are normalized by the total
number of events in the sample rather than the area of each individual
plot.  (a) The \costhv{} distribution for \qsq{}/$\qsq{}_{\rm max} <
0.5$ (b) The \costhv{} distribution for \qsq{}/$\qsq{}_{\rm max} >
0.5$ (c) The \costhl{} distribution for \qsq{}/$\qsq{}_{\rm max} <
0.5$ (d) The \costhl{} distribution for \qsq{}/$\qsq{}_{\rm max} >
0.5$
\label{CVCL}}
\end{figure}

\begin{figure}[htp]
\includegraphics[height=2.in]{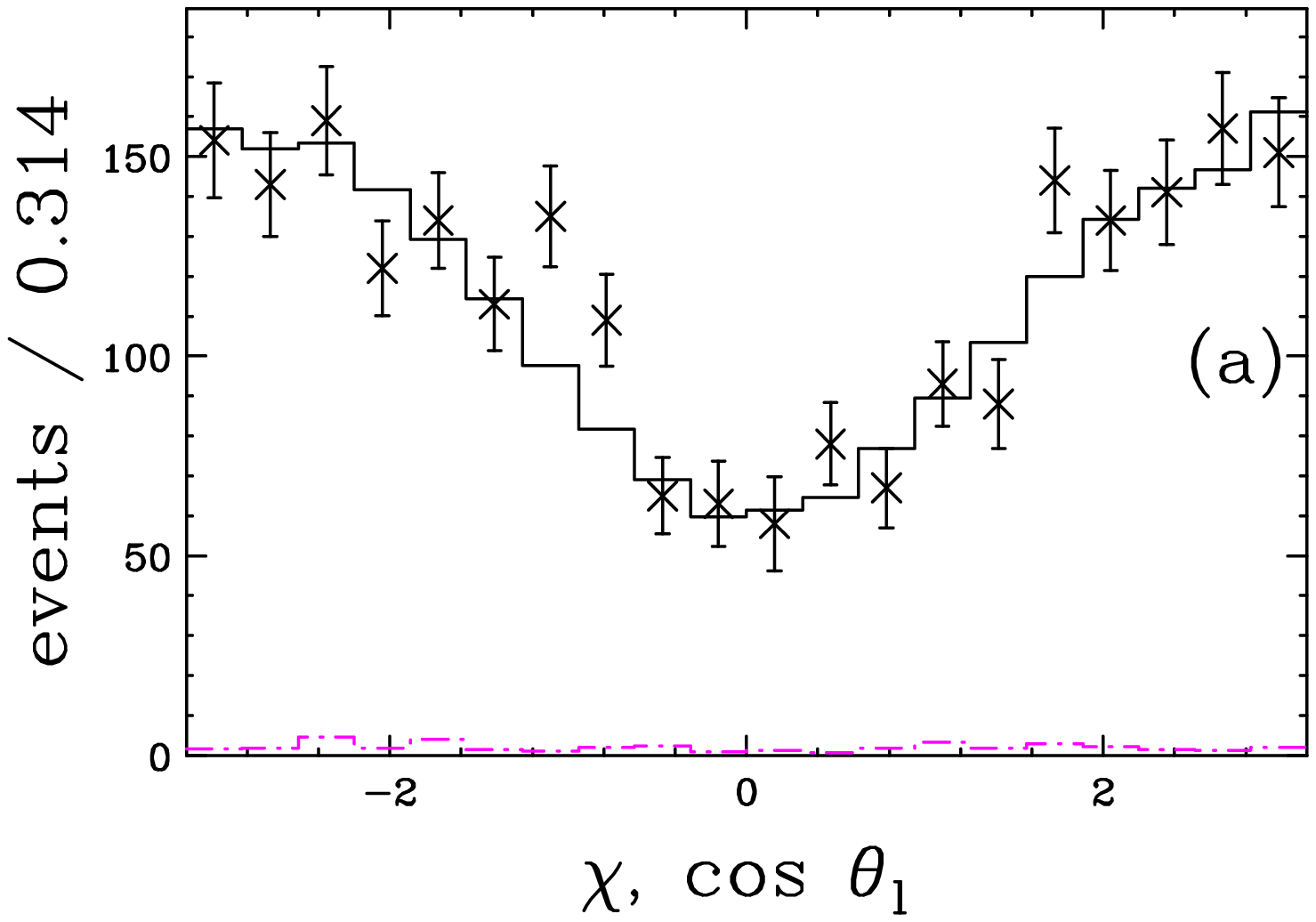}
\includegraphics[height=2.in]{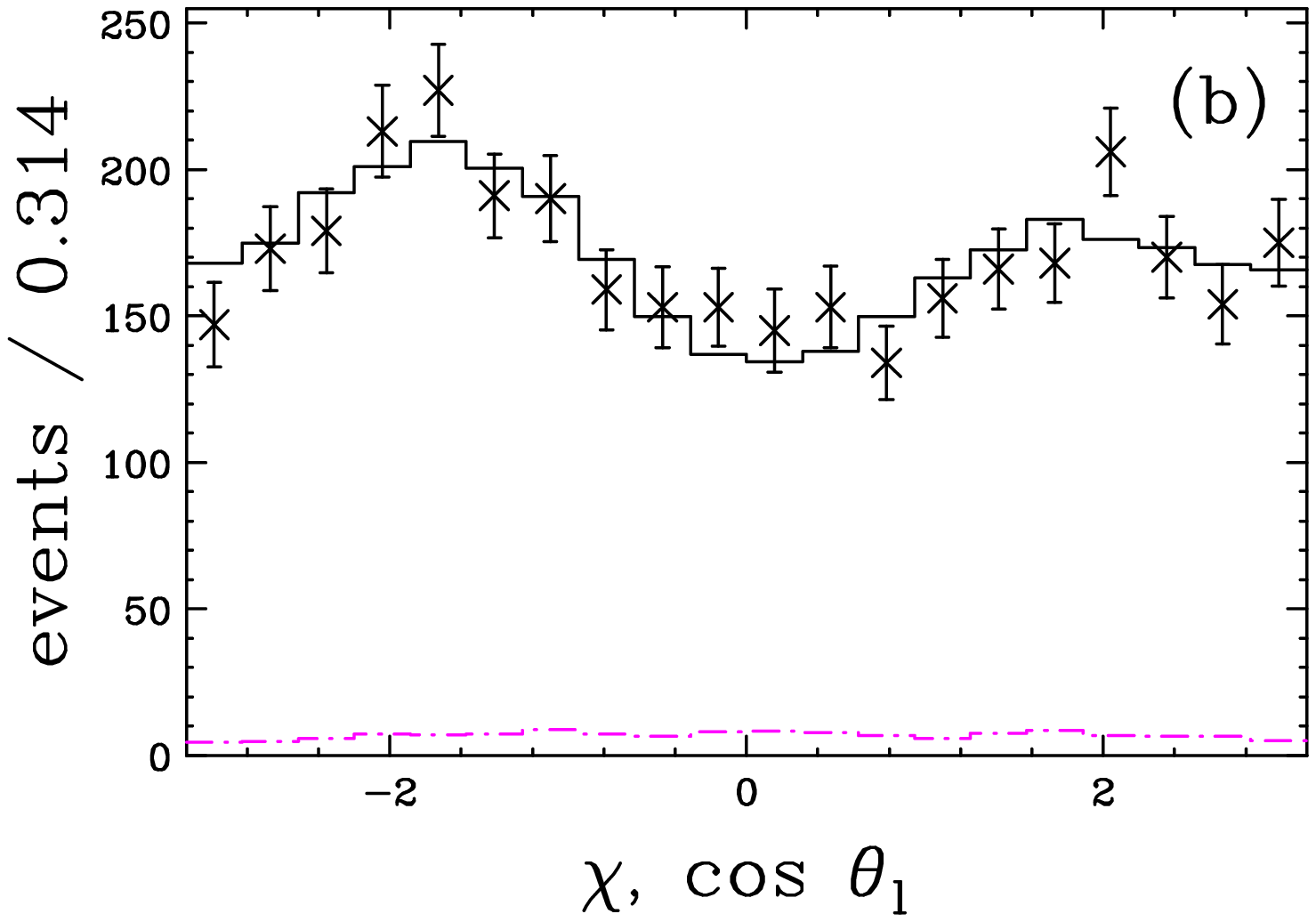}
\includegraphics[height=2.in]{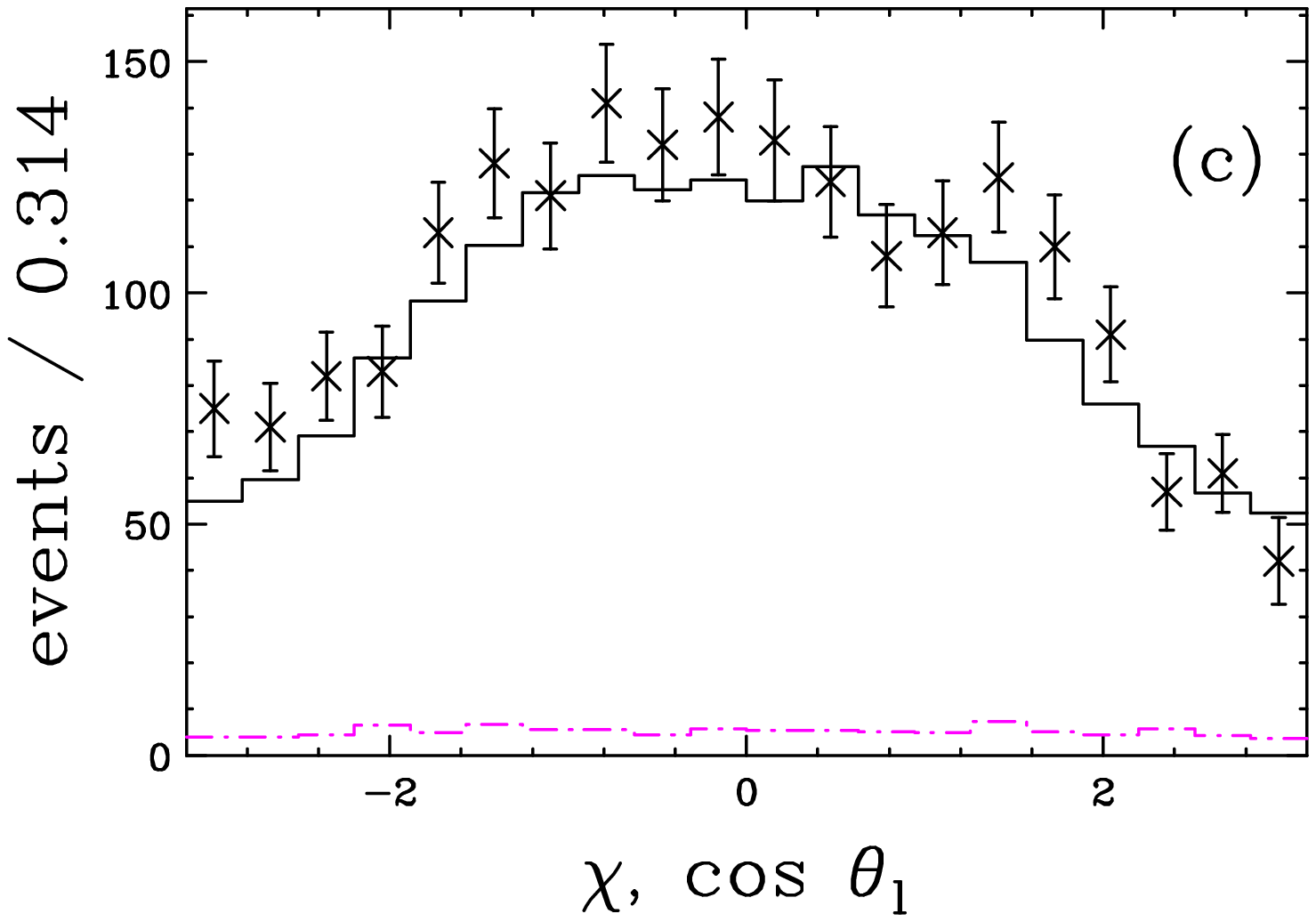}
\includegraphics[height=2.in]{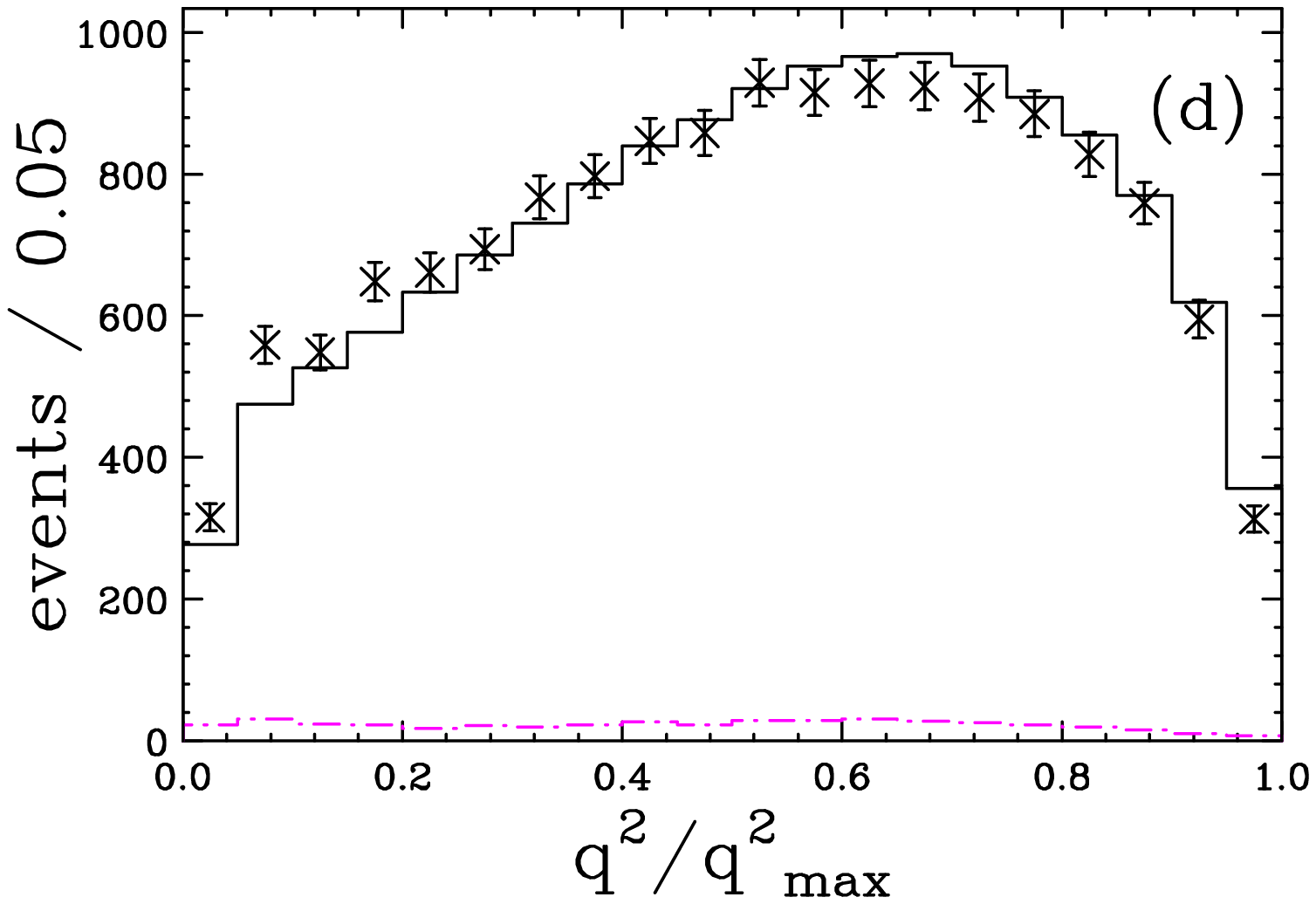}
\caption{Acoplanarity projections with $\costhl{} < 0$ and
three ranges of \costhv{}.  (a) $\costhv{} < -0.5$ (b) $-0.5 <
\costhv{} < 0.5$ and (c) $0.5 < \costhv{}$. (d) The $\qsq/\qsq_{\rm max}$
projection. The data are the points with error bars. The
models are the solid line histograms. 
The predicted charm backgrounds are the dashed lines.
\label{acop}}
\end{figure}

Figures \ref{CVCL} and \ref{acop} compare the data and model for
several of the more interesting projections of \costhv{}, \costhl{},
$\chi$ and \qsq{}. No \qsq{} cut is applied in these projections. Most
of these distributions follow the predicted values reasonably well
with the exception of the low $\qsq{}$ \costhl{} projection (Figure
\ref{CVCL} (c)) for the reasons discussed above.


The expected relative amounts of the $\cos{\chi}$ and $\cos{2 \chi}$
contributions and their phase variation as a function of \costhv{} is
well reproduced by our data as is the $\chi \leftrightarrow -\chi$
asymmetry created by the $s$-wave interference. The respective
$\costhl{} > 0$ projections in data are also well matched by the model
but show less variation than the acoplanarity distributions shown in
Figure \ref{acop}.

Our $s$-wave amplitude fit produced an amplitude modulus of $ A = 0.330
\pm 0.022 \pm 0.015~\rm{GeV}^{-1}$ and a phase of $\delta = 0.68 \pm 0.07
\pm 0.05~{\rm rad}$. Our estimate of the $s$-wave systematic error was
based on the sample variance over 35 fits run with different analysis
cuts. We varied such cuts as the particle identification cuts,
vertexing cuts, and visible mass and energy cuts.

This result
was then fed into our form factor fit to produce our \rvee{} and
\rtwo{} measurements values.

A far more extensive systematic error analysis was made for the form
factor analysis since these are actual physical parameters rather than
an effective description of an interfering amplitude which is only 
validated in the vicinity of the \krzb{} pole.

\section{Form Factor Ratio Systematic Errors}
Three basic approaches were used to determine the systematic error on
the form factor ratios. In the first approach, we measured the stability of
the branching ratio with respect to variations in analysis cuts
designed to suppress backgrounds. In these studies we varied cuts such
as the detachment criteria, the secondary vertex quality, the minimal
number of tracks in our primary vertex, particle identification cuts,
visible momenta cuts, etc. Fifteen such cut sets were considered.  In
the second approach, we split our sample according to a variety of
criteria deemed relevant to our acceptance, production, and decay
models and estimated a systematic based on the consistency of the form
factor ratio measurements among the split samples. We split our sample based
on the visible $D^+$ momentum, particle versus antiparticle, and
whether or not the \mkpi{} mass was above or below 0.9~\gevcsq{}. This
later split was based on our previous observation~\cite{anomaly} of a
large \costhv{} asymmetry that developed for events with $\mkpi{} <
0.9~\gevcsq{}$ due to the $s$-wave amplitude interference.  In the third
approach we checked the stability of the branching fraction as we
varied specific parameters in our Monte Carlo model and fitting
procedure.  These included varying the level of the charm background
Monte Carlo, and the value of the $r_3$ form factor ratio as a uniform
variable over the range $-2 < r_3 < 2$.  

Leaving out the $s$-wave amplitude contribution in our form factor fits
{\it entirely} shifted both $r_v$ and $r_2$ downward by only 
$0.5 \sigma$. Given the insensitivity of our form factor fits to the
$s$-wave amplitude, no systematic error was assessed for uncertainty in
the $s$-wave parameters.  Combining all three non-zero systematic error
estimates in quadrature we find \rvresult{} and \rtworesult{}.

\mysection{Summary}

We presented a fit of the $s$-wave amplitude. We obtained an amplitude
modulus of $ A = 0.330 \pm 0.022 \pm 0.015~{\rm GeV}^{-1}$ and a phase
of $\delta = 0.68 \pm 0.07 \pm 0.05~{\rm rad}$ in reasonable agreement
with our very informal, previous~\cite{anomaly} estimate of $A = 0.36
\exp(i \pi/4)~{\rm GeV}^{-1}$.  The inclusion of the $s$-wave amplitude
dramatically improved the the quality of our form factor fits but
created only minor shifts in the resulting form factor ratio values.
\begin{table}[htp]
\caption{Measurements of the \krzlndk{} form factor ratios}
\begin{center}
\begin{tabular}{l|l|l}
Group & \rvee{} & \rtwo \\
\hline \hline
This work & \rvvalue{}  & \rtwovalue{} \\
BEATRICE \cite{beatrice} & $1.45 \pm 0.23 \pm 0.07$ & $1.00 \pm 0.15 \pm 0.03$ \\
E791 (e) \cite{e791e}  & $1.90 \pm 0.11 \pm 0.09$ & $0.71 \pm 0.08 \pm 0.09$ \\
E791 ($\mu$) \cite{e791mu} & $1.84 \pm 0.11 \pm0.09 $ & $0.75 \pm0.08 \pm 0.09 $ \\
E687 \cite{e687} & $1.74 \pm 0.27 \pm 0.28 $ & $0.78 \pm 0.18 \pm 0.11 $ \\
E653 \cite{e653} & $2.00 \pm 0.33 \pm 0.16 $ & $0.82 \pm 0.22 \pm 0.11 $ \\
E691 \cite{e691} & $2.0 \pm 0.6 \pm 0.3 $ & $0.0 \pm 0.5 \pm 0.2 $ \\
\end{tabular}
\end{center}
\label{ff_table}
\end{table}

Table \ref{ff_table} summarizes measurements of the \rvee{} and
\rtwo{} form factor ratios.  Our measurement is the first one to include the
effects on the acceptance due to changes in the decay angular
distribution brought about by the $s$-wave interference.  We are
consistent with the most recent previous measurement by the BEATRICE
Collaboration.  Our \rvee{} value is about 2.9 standard deviations
below the average of the two (previously most precise) measurements by
the E791 Collaboration although consistent with their value of
\rtwo{}.

\mysection{Acknowlegments}

We wish to acknowledge the assistance of the staffs of Fermi National
Accelerator Laboratory, the INFN of Italy, and the physics departments
of the collaborating institutions. This research was supported in part
by the U.~S.  National Science Foundation, the U.~S. Department of
Energy, the Italian Istituto Nazionale di Fisica Nucleare and
Ministero dell'Universit\`a e della Ricerca Scientifica e Tecnologica,
the Brazilian Conselho Nacional de Desenvolvimento Cient\'{\i}fico e
Tecnol\'ogico, CONACyT-M\'exico, the Korean Ministry of Education, and
the Korean Science and Engineering Foundation.

\end{document}